\documentclass{vgtc}

\graphicspath{{figures/}{data/}{./}}

\usepackage{times}

\usepackage{booktabs}
\usepackage{mathptmx}
\usepackage{amsmath}
\usepackage{amssymb}

\onlineid{1160}
\vgtccategory{Analytics \& Decisions} 
\vgtcinsertpkg

\title{Drag, Infer, Reproject: Grounding LLMs through Spatial Interaction for Image Clustering}

\author{Yang Liu\thanks{e-mail: yangliu07@vt.edu}\\ %
        \scriptsize Virginia Tech %
\and Xuxin Tang\thanks{e-mail: xuxintang@vt.edu}\\ %
     \scriptsize Virginia Tech %
\and Jiahao Xu\thanks{e-mail: jiahao@vt.edu}\\ %
     \scriptsize Virginia Tech
\and Chris North\thanks{e-mail: north@vt.edu}\\ %
     \scriptsize Virginia Tech}

\teaser{
  \centering
  \includegraphics[width=\textwidth]{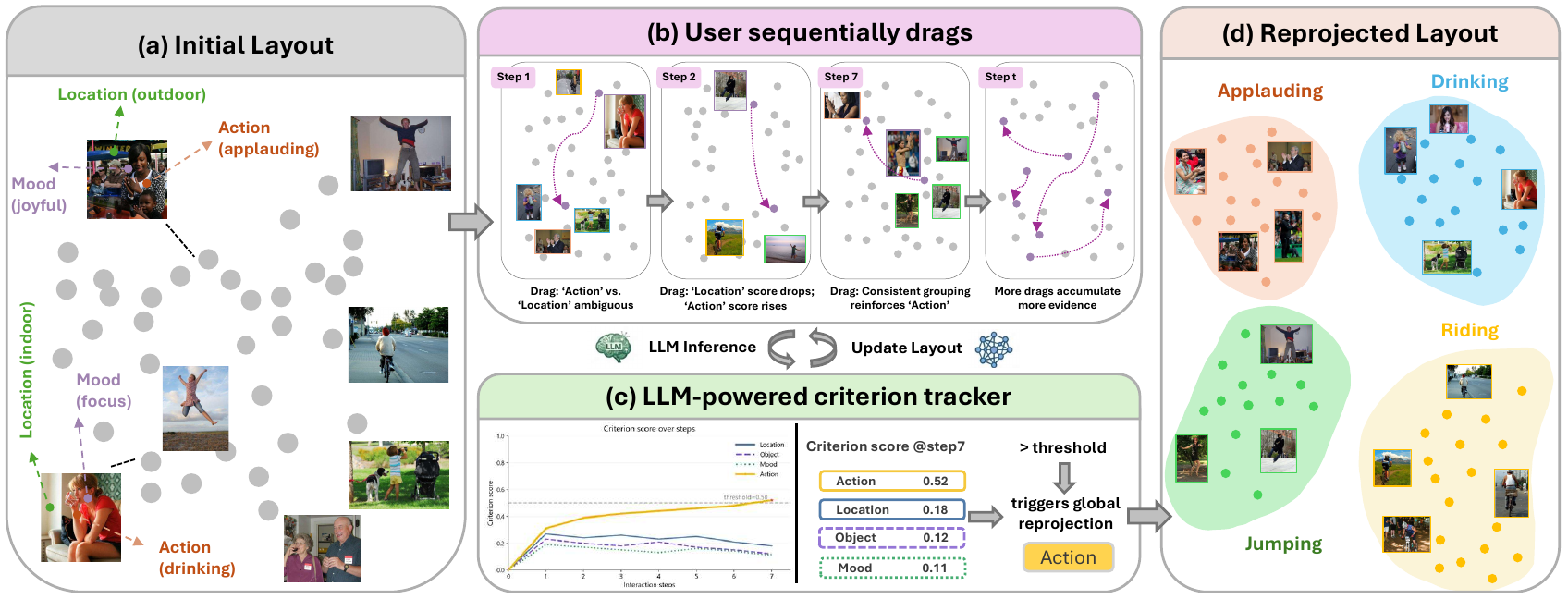}
  \caption{CriterionSI infers and refines the user's clustering criterion from incremental drag interactions. (a)~An initial EVA-CLIP projection shows images in a semantically mixed layout. (b)~The user drags images; the tracker considers multiple candidate criteria. (c)~Once consistent evidence accumulates, the tracker commits to a criterion and assigns values to all images via LLMs. (d)~Criterion-guided reprojection reshapes the entire layout, placing images into semantically coherent clusters.}
  \label{fig:teaser}
}

\abstract{
Dimension reduction and semantic interaction support image clustering by making similarity structure visible and manipulable. Existing semantic interaction methods encode users' clustering criterion (a user-interpretable semantic dimension, e.g., \textit{action}, \textit{location}, or \textit{mood}) from direct manipulation to steer reprojection, giving users direct control over the resulting layout. Yet they typically depend on learned embeddings or a predefined criterion. In practice, users' clustering criterion often emerges gradually and becomes refined through interaction rather than being fully clear at the outset. In this work, we present \textbf{CriterionSI} (\textbf{Criterion}-guided \textbf{S}emantic \textbf{I}nteraction), a method that translates incremental drag interactions into criterion-guided reprojection. CriterionSI uses large language models to infer and refine the clustering criterion from sequential user drags, while grounding semantic interpretation in human-provided feedback rather than fixed prior assumptions. CriterionSI combines the inferred criterion with local drags to guide global reprojection. The simulation-based evaluation and usage scenario demonstrate that CriterionSI can discover and refine the target criterion from sequential interactions and progressively produce criterion-aligned clustering layouts. Our code and data are available at: \url{https://github.com/4C79/CriterionSI}.
}


\begin{document}

\firstsection{Introduction}
\maketitle

Visual analytics enables users to gain insight into complex data through interactive visual exploration~\cite{Keim2008VisualAnalytics,Endert2012SemanticInteraction}. 
In this context, image analysis often involves clustering collections of images to discover similarity relationships and identify meaningful structure in high-dimensional data~\cite{Jain2010DataClustering, kwon2017clustervision}. 
Dimension reduction (DR) supports this process by mapping high-dimensional embeddings into 2D spaces, where spatial proximity serves as a perceptual cue for grouping and pattern discovery~\cite{ware2019information, nonato2018multidimensional, espadoto2019toward}.

In practice, users rarely begin with a fully specified clustering criterion~\cite{Sacha2014KnowledgeGeneration,Pirolli2005Sensemaking}. Instead, they express partial and evolving intent through sequential interactions with the projection~\cite{sacha2016visual, cavallo2018clustrophile, xia2021revisiting}. Each individual manipulation, such as dragging an image toward a group, may admit multiple plausible semantic interpretations, making single-step reasoning inherently ambiguous~\cite{Wenskovitch2024Ambiguous}. As a result, the clustering criterion becomes identifiable through the accumulation and refinement of evidence across interaction steps.

Recent advances in vision-language embeddings and Large Language Models (LLMs) provide rich visual-semantic embeddings and strong capabilities in semantic reasoning, making them effective for interpreting and organizing high-level features~\cite{zhang2024vision}. Building on these capabilities, recent work has explored using LLMs to steer data clustering and visual representations, for example, by incorporating textual criteria or guiding model behavior through prompts and interaction~\cite{Kwon2024ICTC, Oliveira2025UserSteerable, Shao2024LEVA}. In parallel, deep learning-based methods extend semantic interaction (SI) to neural models~\cite{Bian2021DeepSI, Bian2024NeuralSI}, including image projections that learn updated embedding weights from user demonstrations~\cite{Lin2024ImageSI, wei2024spaceediting}, though the organizing criterion remains implicit in the learned representation. This gap motivates making the clustering criterion explicit and reasoning about it with LLMs while preserving direct spatial interaction.

In this work, we present CriterionSI, a method for criterion-guided image clustering from sequential user interactions, as shown in \cref{fig:teaser}. Here, a clustering criterion refers to any semantic dimension that organizes images into groups. CriterionSI treats each drag as partial evidence of an evolving clustering criterion and uses an LLM-powered criterion tracker to infer, verify, and stabilize candidate criteria over interaction steps. Once the criterion stabilizes, CriterionSI assigns semantic values to all images with LLMs and combines the resulting criterion constraints with visual similarity and drag constraints to guide global reprojection, aligning the layout with the user's clustering intent. Our contributions are: (1)~An LLM-powered criterion tracker that infers clustering criterion from sequential drag interactions. (2)~A criterion-guided reprojection pipeline that propagates the inferred criteria to global layout updates. (3)~A simulation-based evaluation demonstrating criterion stabilization and layout improvement over existing methods.

\section{Related work}

\textbf{Interactive projection and semantic interaction.}
DR techniques project high-dimensional data into 2D layouts~\cite{Jain2010DataClustering} that users can inspect and manipulate. SI~\cite{Endert2012SemanticInteraction,Wenskovitch2024Ambiguous} treats direct manipulations of the projection as implicit expressions of analytic intent, which a backing model translates into layout updates. Andromeda~\cite{Self2018Andromeda} adjusts per-dimension weights in Weighted Multi-Dimensional Scaling (WMDS) from SI. DeepSI~\cite{Bian2021DeepSI} fine-tunes text embeddings within the interaction loop, and ImageSI~\cite{Lin2024ImageSI} extends this paradigm to images by learning embedding weights through inverse MDS on image features. Related work on latent-space editing for images~\cite{wei2024spaceediting} similarly updates representations from user demonstrations. These methods progressively adapt how projections respond to SI, yet the clustering criterion remains implicit in the updated model: the system never surfaces a named criterion, nor can it generalize the inferred intent to images the user has not manipulated.

\textbf{Language-conditioned image clustering.}
A complementary line of work treats the clustering criterion as an explicit user-provided natural-language input. IC$|$TC~\cite{Kwon2024ICTC} clusters images under user-provided textual criteria using vision-language models. Multi-MaP~\cite{yao2024multi} supports personalized multiple clustering in which users supply brief keywords and LLMs formulate criterion-specific textual contexts. Interactive Semantic Mapping~\cite{Oliveira2025UserSteerable} steers projections through natural-language prompts combined with zero-shot classification. More broadly, LLM-augmented visual analytics workflows~\cite{Shao2024LEVA, tang2026semantic} demonstrate the value of symbolic semantic reasoning in user-facing tools. These approaches give users direct control over \emph{what} to cluster by, but require predefined criteria, limiting their applicability when intent is still being formed.

CriterionSI bridges these gaps, taking only drag actions yet surfacing an explicit, named criterion that emerges through interaction rather than being specified upfront.

\begin{figure*}[t]
\centering
\includegraphics[width=0.85\textwidth]{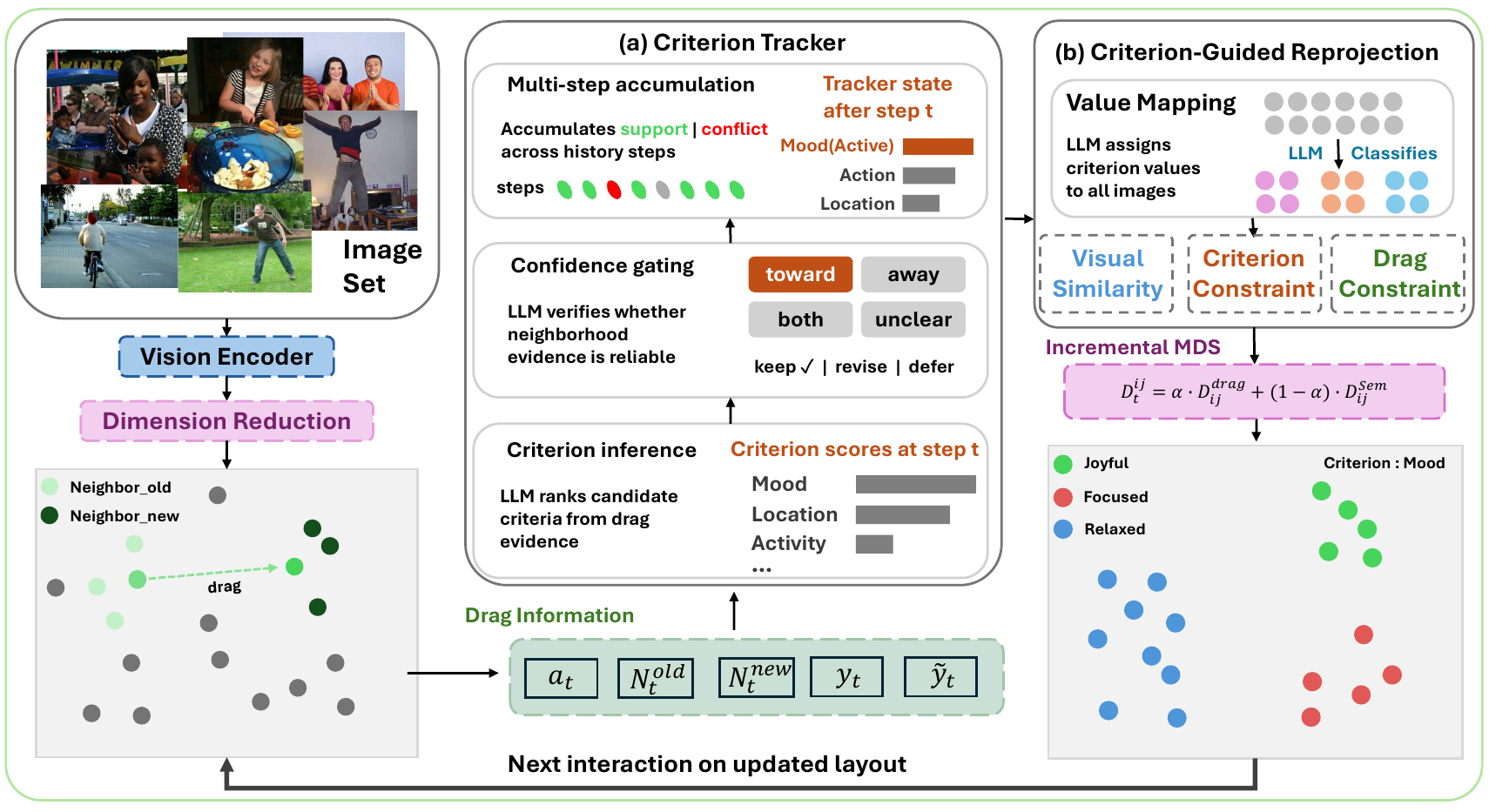}
\caption{Overview of CriterionSI: (a) Criterion Tracker interprets each drag as partial evidence of a latent clustering criterion. (b) Criterion-Guided Reprojection reshapes pairwise distances and updates the layout through incremental MDS, closing the loop for the next interaction.}
\label{fig:pipeline}
\end{figure*}

\section{Method}
\label{sec:method}

CriterionSI targets exploratory analytical settings in which the user starts with a tentative sense of how to cluster the image collection, and refines this clustering criterion progressively through drag interactions. Given a collection of images encoded by a frozen vision encoder and projected to 2D via MDS, CriterionSI obtains an initial layout $Y^{(0)}$ on which all subsequent interactions and updates operate. The method then proceeds in three stages (\cref{fig:pipeline}), realized by two components that form a closed loop.

\subsection{Interaction capture}

At step $t$, the user drags an anchor image $a_t$ 
from one canvas position to another. We encode this 
action as \begin{equation} e_t = (a_t,\; N_t^{\mathrm{old}},\; N_t^{\mathrm{new}},\; y_t,\; \tilde{y}_t), \label{eq:event} \end{equation} where $y_t, \tilde{y}_t$ are the anchor's pre- and post-drag positions and $N_t^{\mathrm{old}}, N_t^{\mathrm{new}}$ are the $k$-nearest neighborhoods around $y_t$ and $\tilde{y}_t$ on the 2D canvas. From the geometry of the trajectory relative to the two neighborhoods, we classify the interaction into a mode $g_t \in \{\textit{toward}, \textit{away}, \textit{both}, \textit{unclear}\}$, where \textit{both} denotes simultaneous toward and away signals. The mode $g_t$ then feeds the criterion inference and confidence gating stages that follow.

\subsection{Criterion-guided Semantic Interaction}
\label{sec:tracker}

\textbf{Criterion tracker.}
The criterion tracker (\cref{fig:pipeline}a) makes at most three LLM calls per step: criterion inference, confidence gating, and conditional value mapping. A deterministic state machine then consolidates these per-step hypotheses into a persistent decision.

\textbf{Criterion inference.}
Given the anchor image, both neighborhoods, and the interaction mode $g_t$, the LLM ranks candidate criteria and returns up to three, denoted $c_1, c_2, c_3$ with associated scores $\gamma_1, \gamma_2, 
\gamma_3 \in [0,1]$ and $\gamma_1 \geq \gamma_2 \geq \gamma_3$. Because $N_t^{\mathrm{new}}$ reflects what the user actively wants to group while $N_t^{\mathrm{old}}$ may merely reflect the previous layout, \textit{toward} evidence is weighted more heavily when the two disagree. A small set of reference dimensions, drawn from dataset metadata or summarized by an LLM from a sample of the collection, serves as a shared vocabulary to reduce criterion fragmentation while preserving open-endedness.

\textbf{Confidence gating.}
When the top candidates are not clearly separated, a verification pass re-examines whether neighborhood evidence is reliable given the interaction mode $g_t$, and outputs one of three decisions: \textit{keep}, \textit{revise}, or \textit{defer}. The gating stage also emits a relation label $r_t \in \{\mathrm{support},~\mathrm{conflict}, ~\mathrm{unclear}\}$ describing how the (possibly revised) top candidate relates to the active criterion, which drives the state transitions below.

\textbf{Multi-step accumulation.}
The state machine maintains the active criterion $c^*$, a support counter $n$, and a running aggregated score $\hat{\gamma}_c$ for each candidate that is reinforced when an interaction supports $c$ and decayed under conflicting or unclear evidence; transitions consult $\hat{\gamma}_c$ rather than the per-step $\gamma_t$ to smooth the uncalibrated per-step scores. The transition rules are:

\begin{itemize}
\setlength{\itemsep}{0pt}
\setlength{\parskip}{1pt} 

\item \textit{Activation}: a candidate becomes the active criterion $c^*$ once it consistently ranks first across the window with $\hat{\gamma}_c \geq \hat{\gamma}_{\min}$.
\item \textit{Reinforcement}: $r_t{=}\mathrm{support}$ increments the support counter $n$ and decays the staged challenger's score $\hat{\gamma}_{c_1}$.
\item \textit{Challenge and switch}: $r_t{=}\mathrm{conflict}$ buffers the top candidate $c_1 \neq c^*$ as a challenger; the challenger replaces $c^*$ only when $\hat{\gamma}_{c_1} - \hat{\gamma}_{c^*} > \delta$ is sustained across the window.
\item \textit{Propagation gate}: when $c^*$ is active and $n$ exceeds a minimum, the LLM is queried to assign each image a value $v_i$ under $c^*$ (\cref{fig:pipeline}b, top); otherwise reprojection updates from drag blending alone, with $D^{\mathrm{sem}} {=} D^{(t-1)}$ unchanged.

\end{itemize}

Once the tracker produces stable value assignments, the system reshapes the layout through criterion-guided reprojection, drag-constrained blending, and incremental MDS (\cref{fig:pipeline}b). All distances below are computed in the 2D canvas coordinate system of the current layout, ensuring that the subsequent blending operates on a shared scale.

\textbf{Criterion-guided reprojection.}
Let $y_i^{(t-1)}$ denote the 2D canvas position of image $i$ after the previous reprojection step, and let $D_{ij}^{(t-1)} = \|y_i^{(t-1)} - y_j^{(t-1)}\|_2$ be the pairwise distance between images $i$ and $j$ in the current layout. For the anchor $i = a_t$, $y_t$ denotes its pre-drag position and $\tilde{y}_t$ denotes its post-drag target. We scale these distances by value agreement under the active criterion:
\begin{equation}
D_{ij}^{\mathrm{sem}} = D_{ij}^{(t-1)} \cdot
\begin{cases}
1 - \beta, & v_i = v_j, \\
1 + \beta, & v_i \neq v_j,
\end{cases}
\label{eq:shaping}
\end{equation}
where $\beta \in [0, 1)$ controls the shaping strength. Same-value pairs are pulled closer and different-value pairs are pushed apart. When no stable criterion is available, $D^{\mathrm{sem}} = D^{(t-1)}$.

\textbf{Drag-constrained blending.}
For pairs involving the drag anchor, we compute $D_{ij}^{\mathrm{drag}} = \|\tilde{y}_{a_t} - y_j^{(t-1)}\|_2$, the distance implied by placing the anchor at its target position while neighbors retain their current coordinates. The target distances for these pairs are:
\begin{equation}
D_{ij}^{\mathrm{target}} = \alpha \cdot D_{ij}^{\mathrm{drag}} 
+ (1-\alpha) \cdot D_{ij}^{\mathrm{sem}}, \quad (i,j) \in P_t,
\label{eq:blending}
\end{equation}
where $\alpha \in [0, 1]$ controls drag weight and $P_t = \{\{a_t, j\} \mid j \in N_t^{\mathrm{old}} \cup N_t^{\mathrm{new}}\}$ is the set of anchor-neighbor pairs. For all other pairs, $D_{ij}^{\mathrm{target}} = D_{ij}^{\mathrm{sem}}$.

\textbf{Incremental MDS.}
The updated layout $Y^{(t)}$ is obtained by minimizing stress $\sum_{i<j}(\|y_i - y_j\| - D_{ij}^{\mathrm{target}})^2$ from the current positions, with the $K \geq 0$ most recent drag anchors pinned at their post-drag targets. This sliding-window pinning ensures that early drags do not dominate the optimization as the interaction progresses.

\section{Usage Scenario}
\label{sec:usage_scenario}

We illustrate how the tracker evolves through a simulated session on the \textit{Mood} task (clustering the Action40 collection by mood). All silhouette values in this section are from that one run, and experiment settings follow~\cref{sec:experiment_setup}.

\begin{figure}[!t]
\centering
\includegraphics[width=\columnwidth]{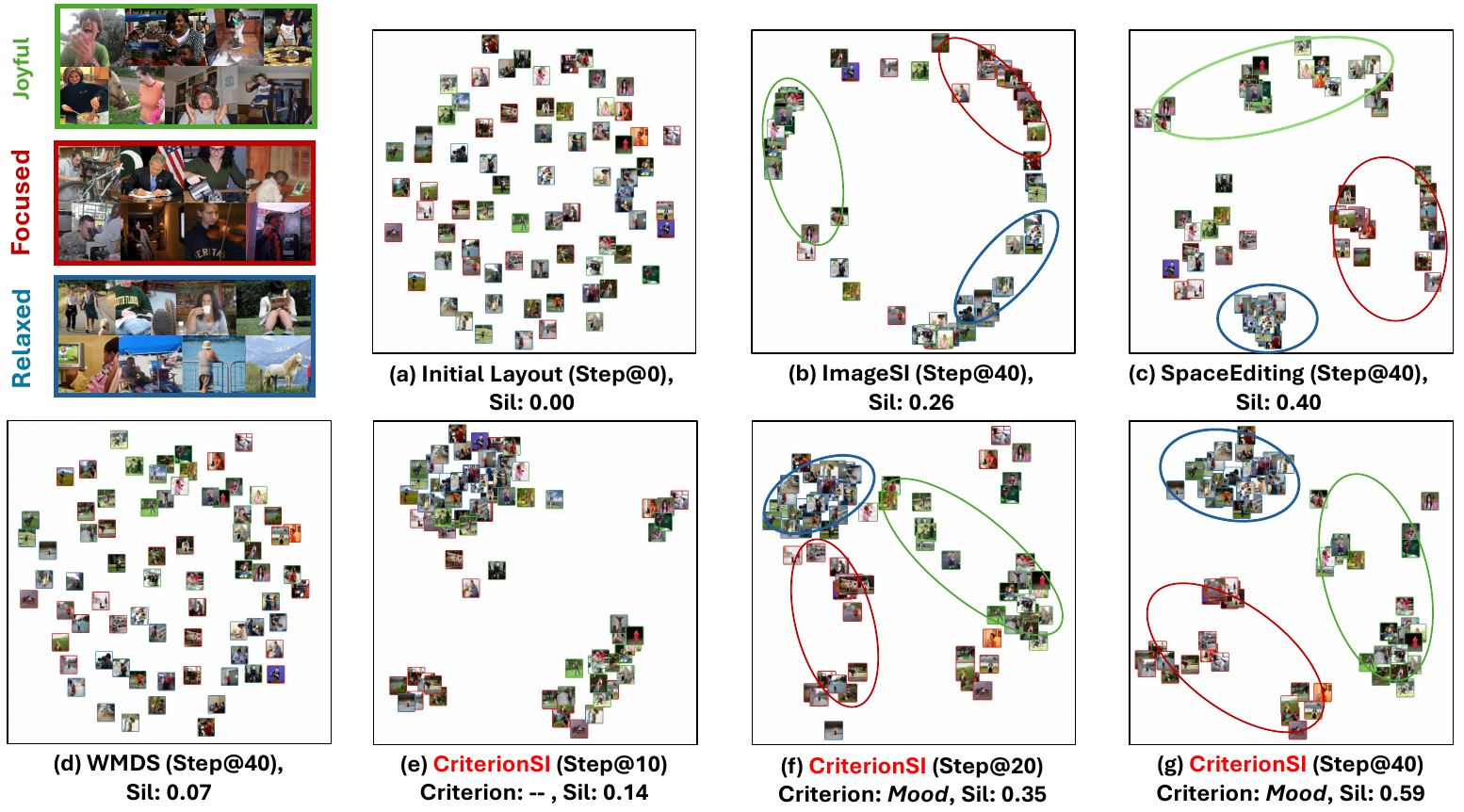}
\caption{DR plots after interaction on the \textit{Mood} task. (same as \cref{sec:usage_scenario})
(a)~Initial layout; (b-d)~ImageSI, SpaceEditing, WMDS at step~40;
(e-g)~CriterionSI~(Gemini~cls) at steps~10, 20, and 40.
}
\label{fig:final_layout}
\end{figure}

\textbf{Initial projection.}
The initial layout is arranged by visual similarity, leaving the three mode classes spatially intermixed (\cref{fig:final_layout}a).

\textbf{Ambiguous early drags (steps 1--5).}
In the first few drags, the LLM returns near-tied candidate criteria across steps. After aggregation over the sliding window, no candidate reaches the activation threshold, so the tracker remains uncommitted. The layout shows no clear cluster structure ($s_5 = -0.03$):
\begin{quote}\small\itshape
``selected\_criterion: \textbf{social context}; 
selected\_value: \textbf{solitary} (0.72); \\ 
alternatives: \textbf{mood}/\textbf{joyful} (0.60), 
\textbf{age}/\textbf{children} (0.60).''\hfill{\rm[step~5]}
\end{quote}

\textbf{Criterion stabilization (step 6).}
The next \textit{toward} drag groups a man-jumping image with a bubble-blowing scene; the LLM disambiguates, and \textit{mood}'s aggregated score $\hat\gamma_{\textit{mood}} = 0.64$ crosses the activation threshold over the sliding window, while the per-step score reaches $0.90$ at this step.
\begin{quote}\small\itshape
``criterion\_action:\textbf{activate}; 
selected\_criterion: \textbf{Mood} (0.90). Anchor and new 
group consistently depict active joy, playfulness, or 
enthusiastic enjoyment.''\hfill{\rm[step~6]}
\end{quote}
Silhouette turns positive at activation ($s_6 = 0.04$) and grows incrementally as the criterion-guided reprojection reshapes the layout over subsequent steps, reaching $s_{20} = 0.35$ 
(\cref{fig:final_layout}f).

\textbf{Refinement under momentary conflict (steps~7--40).}
Supporting \textit{toward} drags re-confirm \textit{mood}; even a step-20 challenge nominating \textit{Age Extremes} (0.95) fails the state machine's switch criterion, since its lead over mood does not persist across the sliding window. At step~28 a bubble-blowing 
anchor briefly prompts the LLM to rank 
\textit{Recreational Activity} above \textit{mood}:
\begin{quote}\small\itshape
``active\_criterion\_check: \textbf{conflict}; 
selected\_criterion: \textbf{Recreational Activity} (0.95).''
\hfill{\rm[step~28]}
\end{quote}
Because the state machine evaluates accumulated support across multiple steps rather than a single peak, this one-step spike fails to meet the switch threshold; the cumulative evidence for \textit{mood} from prior steps still dominates. Value mapping therefore continues under $c^* = \textit{mood}$, and the silhouette holds at $s_{40} = 0.59$ with the three mood clusters cleanly resolved (\cref{fig:final_layout}g).

\section{Evaluation}
\label{sec:evaluation}

We evaluate CriterionSI through a simulation-based quantitative comparison and a representative usage scenario on the \textit{Mood} task.

\textbf{Implementation details.}
Images are encoded with EVA-CLIP ViT-G/14~\cite{Sun2023EVACLIP} and all LLM calls use Gemini~2.5~Flash. We use neighborhood $k{=}3$, window $W{=}3$, activation $\hat{\gamma}_{\min} {=} 0.5$, switch margin $\delta{=}0.15$, shaping $\beta{=}0.3$, drag-blend $\alpha{=}0.7$, and anchor-pinning $K{=}5$. Prompts, datasets, more implementation details, and more usage scenarios are in the supplemental material.

\begin{table}[b]
\centering
\small
\caption{Compared methods by the signals used to update the layout.}
\label{tab:methods}
\begin{tabular}{lcccc}
\toprule
Method & Drag & Visual & Language & Criterion \\
\midrule
WMDS~\cite{Self2018Andromeda} & \checkmark & & & \\
ImageSI~\cite{Lin2024ImageSI} & \checkmark & \checkmark & & \\
SpaceEditing~\cite{wei2024spaceediting} & \checkmark & \checkmark & &\\
\textbf{CriterionSI~(ours)} & \checkmark & \checkmark & \checkmark & inferred \\
ISM~\cite{Oliveira2025UserSteerable} & & \checkmark & \checkmark & predefined \\
\bottomrule
\end{tabular}
\end{table}

\subsection{Experimental Setup}
\label{sec:experiment_setup}

\textbf{Simulation engine.}
We simulate sequential user behavior with a goal-directed drag generator. At each step, an anchor image is selected from the unvisited pool and moved toward the centroid of the same-label images that have already been positioned. The simulator uses ground-truth labels only for drag target generation; the tracker receives only the drag event and neighborhood images, with no label access. No purity constraints are enforced on neighborhoods, so the method is exposed to noisy and ambiguous evidence.

\textbf{Dataset and comparison methods.}
We use a 72-image subset of Action40~\cite{yao2011human} and cluster it into the three Mood classes \textit{Joyful}, \textit{Focused}, and \textit{Relaxed} defined by IC$|$TC~\cite{Kwon2024ICTC}. We focus on Mood because it is the most challenging of the Action40
dimensions: \textit{mood} is weakly aligned with visual-appearance distances.
All methods share the same 
EVA-CLIP~\cite{Sun2023EVACLIP} initial representation and projection; they differ only in the signals used to update the layout, summarized in \cref{tab:methods}. ISM~\cite{Oliveira2025UserSteerable} refers to the same CriterionSI pipeline without drag interaction: it receives the ground-truth criterion upfront and shares oracle value assignments with CriterionSI~(oracle~cls). The resulting comparison isolates drag-driven criterion tracking and reprojection. We additionally report two variants of our method:
CriterionSI~(Gemini~cls) is the fully automatic pipeline in which Gemini~2.5~Flash performs both criterion inference and value classification; CriterionSI~(oracle~cls) uses ground-truth value assignments to isolate criterion-tracking quality from classification noise.

\begin{figure}[t]
\centering
\includegraphics[width=\columnwidth]{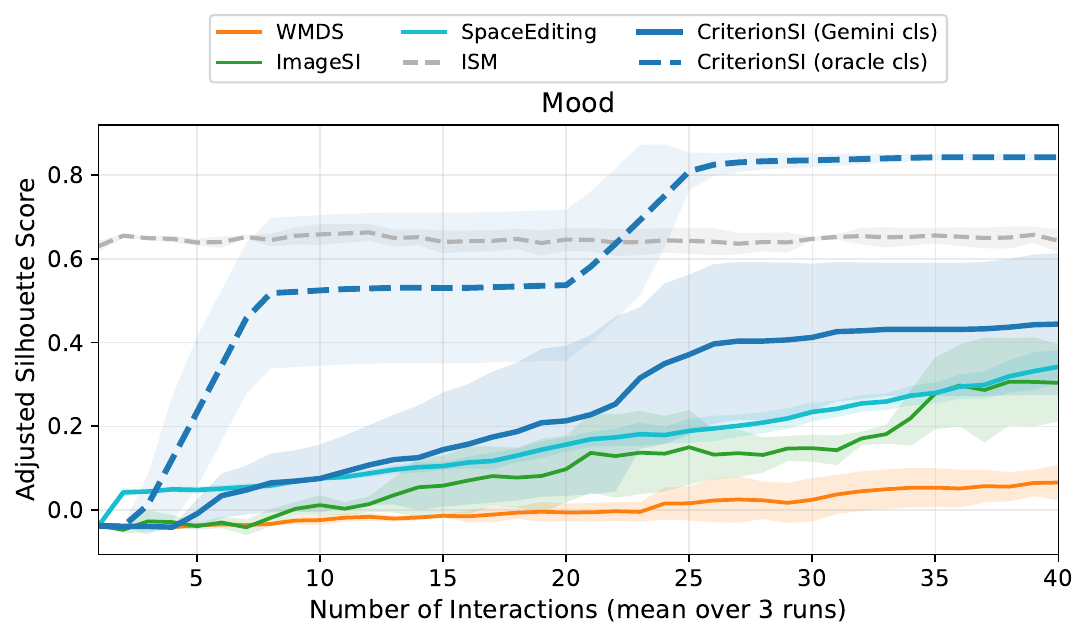}
\caption{Per-step adjusted Silhouette scores on the  \textit{Mood} task; shaded bands show variance across simulated sessions.}
\label{fig:results}
\end{figure}

\subsection{Quantitative Results}
\label{sec:quantitative_results}
\cref{fig:results} shows the per-step silhouette trajectory; all numbers below are averaged over three runs. Starting from $s_0 \approx 0.00$, CriterionSI~(Gemini~cls) reaches $s_{40} = 0.44$,  substantially outperforming other baselines under the same interaction budget. Across our tasks, the criterion commits within roughly 10\% of the dataset size, after which value mapping propagates it to all images at once. WMDS struggles to rise above $s\approx 0.1$, since its reweighting can only reshape visual similarity and cannot separate images that look alike but belong to different moods. ImageSI and SpaceEditing steadily improve but level off in the mid-range, because their weight adaptations struggle to learn discriminative features from drag interactions when the clustering criterion is not visually salient.

Notably, CriterionSI~(oracle~cls) reaches $s_{40} = 0.84$, exceeding ISM ($s_{40} = 0.65$) by $0.19$. Since the value mapping is controlled, this gap shows that drag-driven tracking and reprojection contribute a local grouping signal that pure criterion-based clustering cannot provide. The Gemini cls shortfall stems from classification noise on \textit{mood} labels, leaving classification fidelity as the bottleneck.

\section{Discussion}

\textbf{Rethinking ``correct'' criterion recovery in SI.}
CriterionSI does not always converge on the exact target label. When the target is \textit{Action}, it may stabilize on \textit{body posture}, since dynamic movements and static postures yield a largely compatible partition. This is not an isolated case: semantically related criteria such as posture and action, or mood and expression, often carve the data along near-parallel boundaries, making criterion tracking a many-to-one problem at the partition level. We observed no clear performance degradation under near-miss convergence; surfacing a neighboring criterion can be informative, revealing semantic dimensions the user had not explicitly considered. We therefore view CriterionSI less as a system that must guess the right label, and more as one that commits to a coherent partition the user can accept, reject, or refine.

\textbf{Co-constructed intent through divergent reasoning.}
Across our sessions, LLMs often propose criteria the user had not articulated, and occasionally reason along dimensions that do not match the user's initial target. Rather than treating these divergences as inference failures, CriterionSI treats them as hypotheses to be tested against continued interaction. Partial human signals accumulate through drags, LLM hypotheses that fail to find support decay, and the criterion that stabilizes is one both parties have implicitly agreed on through different modes of reasoning. Intent is thus neither extracted from the user nor generated by the model, but co-constructed: the state machine is where human evidence and LLM hypotheses negotiate, and the resulting criterion is often one neither party would have written down at step one.

\textbf{Limitations.} Because following a criterion requires a fraction of the images(\cref{sec:quantitative_results}), interaction cost grows with collection size and dataset complexity; proactively proposing the criterion once enough evidence has accumulated could curb this. The evaluation also relies on a simulated, goal-consistent user that excludes human error and mid-session reconsideration; whether these results hold for real users is an open question we leave to future work.

\textbf{Conclusion and Future Work.} 
\label{sec:conclusion} 
In this work, we introduced CriterionSI, a drag-based approach that couples user interaction with LLM-driven criterion inference for image clustering. Simulation experiments showed CriterionSI outperforms prior interactive baselines and rivals settings in which the criterion is specified upfront. For future work, we see CriterionSI as a step toward agentic visualization that reasons alongside users, with proactive criterion suggestion, multi-agent negotiation, and persistent cross-session memory as natural next directions.

\bibliographystyle{abbrv-doi}
\bibliography{template}

\end{document}